\documentclass[floatfix,showkeys,twocolumn,showpacs,preprintnumbers]{revtex4}

\usepackage{graphicx}

\newcommand\lsim{\mathrel{\rlap{\lower4pt\hbox{\hskip1pt$\sim$}}\raise1pt\hbox{$<$}}}
\newcommand\gsim{\mathrel{\rlap{\lower4pt\hbox{\hskip1pt$\sim$}}\raise1pt\hbox{$>$}}}

\begin{document}

\title{
  Effects of food web construction by evolution or immigration
}

\author{Craig R. Powell}
\email{craig@powell.name}
\affiliation{
  Theoretical Physics Group,
  School of Physics and Astronomy,
  University of Manchester,
  Manchester,
  M13 9PL,
  UK
}

\author{Alan J. McKane}
\email{alan.mckane@manchester.ac.uk}
\affiliation{
  Theoretical Physics Group,
  School of Physics and Astronomy,
  University of Manchester,
  Manchester,
  M13 9PL,
  UK
}

\begin{abstract}
We present results contrasting food webs constructed using the same
model where the source of species was either evolution or immigration
from a previously evolved species pool. The overall structure of the
webs are remarkably similar, although we find some important
differences which mainly relate to the percentage of basal and top
species. Food webs assembled from evolved webs also show distinct
plateaux in the number of tropic levels as the resources available to
system increase, in contrast to evolved webs. By equating the
resources available to basal species to area, we are able to examine
the species-area curve created by each process separately. They are
found to correspond to different regimes of the tri-phasic
species-area curve.
\end{abstract}

\pacs{87.23.Cc}

\keywords{Species area relation, Food webs, Trophic levels, Individual-based model}

\maketitle

\section{Introduction}
Models of food web structure fall into several distinct classes.
Early models tended to be either static, where links were assigned
between species according to some rule, or dynamic, but where the
dynamics consisted of population dynamics on a random network.
Examples of the former are the cascade \citep{coh85,coh90} and niche
\citep{wil00} models, and of the latter the work of
\cite{may72,may74}.  More recent approaches have incorporated longer
time scales, allowing for the introduction of new species through
immigration or speciation and for species extinction.  This allows the
web structure to build up over time; the structure of the web emerges
rather than being put in by hand.  The two ways of doing this have
been through assembly models, which introduce new species into the
community from a species pool \citep{law99}, and evolutionary models,
which introduce new species through modification of existing species
-- speciation \citep{dro03}.  The purpose of this paper is to unify
these two approaches by constructing a species pool through an
evolutionary dynamics and then using this to assemble communities.

Assembly and evolutionary models both have two, separated,
time-scales.  On the ecological time-scale the population sizes of the
species in the community change according to the equations of the
population dynamics until they eventually reach a fixed point or some
other attractor.  On a longer time-scale species are introduced
through immigration (assembly models) or by speciation (evolutionary
models).  The new species then competes with the existing species
following the equations of the population dynamics until the system
again reaches equilibrium.  If the new species does not immediately go
extinct, it may either coexist with those species present at its
introduction, or may cause one or more extinctions, potentially
resulting in its own extinction.  The population dynamics takes the
form of differential equations for the population numbers (using, for
instance, Lotka-Volterra, Holling type II, or ratio-dependent
functional responses) and so the extinction threshold has to be
specified.  Typically this is set to be such that if the population of
any particular species falls below 1, it is deemed to be extinct.

Assembly models usually consist of species pools of tens of species
which are labelled as ``plants'', ``herbivores'', ``carnivores'', etc.
The interactions between these various trophic levels are typically
assigned by some rule with a large amount of randomness built in, but
body-size considerations may also be used to decide the predator-prey
relationships.  Early work used numerical integration of
Lotka-Volterra equations, combined with the criterion of local
stability \citep{pos83,dra88,dra90}, although this had some problems
\citep{mor96}, and other methods of deciding whether a particular
community is stable have been used \citep{law96,mor97}.  Species
assembly models are capable of generating reasonably-sized food webs
\citep{law99} through immigration, although the species pool is made
up of species which are randomly assigned rather than having
co-evolved, and as such it is very artificial.  Moreover, only very
simple population dynamics have been investigated, underlining the
overall lack of realism of this approach.

Evolutionary models have been developed during the last decade or so.
Species and their interactions may be specified by traits which define
phenotypic or behavioural characteristics
\citep{cal98,dro01,yos03a,ros06a}, or the strength of interaction
between species may be described through matrices \citep{las01,kon03}.
Additional mechanisms, such as adaptive foraging, may be included and
may add stability to the system \citep{kon03}.  The evolutionary
approach starts from a small number of species and, through the
modification of existing species, is capable of generating large webs.
The choice of population dynamics seems to be important \citep{dro04},
with the simplest types of population dynamics, such as
Lotka-Volterra, being unable to lead to communities with large stable
webs.  A disadvantage of evolutionary models is that it is not clear
what community is being constructed; it seems to be one in which
immigration has played no part.  A more consistent viewpoint would be
to use the community constructed through the evolutionary approach as
a species pool in the sense of the assembly model, and then use these
co-evolved species to create food webs which would be more analogous
to those for which data is collected.  This will be the point of view
we adopt here.

The idea of using a pool of species as a source of immigrants for
colonisation goes back many years and was the central feature of the
theory of island biogeography developed in the 1960s
\citep{pre62,mac63,mac67,sim74,pie79}.  There the pool was called the
``mainland'' and the community of interest the ``island'', and this is
the terminology we will adopt here.  That theory was an equilibrium
theory; it assumed that immigration and local extinctions were in
balance, although with continual overturn of species.  Although there
have been calls for this theory to be updated and extended
\citep{lom00,bro00}, there has been little work trying to do so using
recent developments in modelling tools and techniques.  A notable
exception is the work of Hubbell \citep{hub01} which uses the
mainland/island picture to formulate a neutral theory of biogeography
and biodiversity.  One way to view the work we describe here is as a
generalisation of these ideas to predator-prey interactions,
incorporating many other aspects and leading to food webs with several
trophic levels.

In the following section we reiterate the Webworld mathematical model
presented in \cite{dro01}, which was used to generate the simulation
results presented in this paper.  In previous papers
\citep{dro01,qui05} this model has been used to construct food webs
through evolution of the species present.  We discuss the
modifications used to study the effect of food web construction
through immigration.  In our main results section we examine typical
measures of food webs structure in terms of the resources available to
the system, and in terms of the number of species present.  We then
focus on the species-area relation as a composite measure of food web
behaviour, and examine our results in terms of power-law fitting.  We
conclude with a discussion of the results we have obtained and
possible directions for future work.

\section{Model}
The Webworld model was introduced by \cite{cal98} to link ecological
modelling of food web structure with evolutionary dynamics of species
creation.  A refinement to the population dynamics was introduced by
\cite{dro01}, and it is this model, described in detail below, which
we adopt.  The long-term behaviour of the model, identified by
\cite{dro01}, is for a continual overturn of species to occur with
relatively well defined mean values of such quantities as the number
of species present.

The Webworld model constructs food webs from species defined by a set,
$A$, of ten different attributes which represent phenotypic and
behavioural characteristics pertaining to survival.  Initially an
antisymmetric matrix $m$ is randomly generated to indicate the
relative score of pairs of attributes; the relative score of two
species is defined by
\begin{equation}
  S_{ij}=\frac1L{\rm max}\!\left(0,\sum_{\alpha\in A_i}\sum_{\beta\in A_j}m_
{\alpha\beta}\right),
\end{equation}
where $L=10$ is chosen to give scores $S_{ij}\sim1$.  If $S_{ij}>0$
then species $i$ is capable of feeding on species $j$.  The utility of
this system of defining species is that incremental evolution can
occur by changing one attribute of a member of a selected species to
form a new species, whose scores will be similar to those of the
parent species.  Species are numbered such that $1\le i\le S$, where
$S$ is the number of species present.  Each species has population
$N_i$ subject to the population dynamics.  One special species is
created, denoted as species zero, whose population is fixed at
$N_0=R$.  This species represents the environment, the basic food
source supplying the whole food web.  The value $R$ is the effective
population of the environmental resources, which provides a persistent
food source for the food web as a whole.  A value of $R=10^6$ was
chosen to grow a source community with approximately 100 species, and
hence this value of $R$ is an approximate upper bound for models
utilising this source community.  The maximum value of $R$ used for
evolving communities was prescribed by computational resources.

In the evolutionary model, the first species is created with a random
set of attributes such that its score against the environment species
is non-zero.  Were this not the case, the first species would go
extinct, being unable to feed.  Subsequent species are introduced by
taking one extant species as the parent, and altering one attribute to
create a daughter species of population 1.  No attribute is allowed to
repeat within a single species.  The food web is constructed by
repetition of this speciation mechanism, with extinctions determined
by the population dynamics described below.  The decision to introduce
new species with population 1 is arbitrary except that this is the
smallest population which does not lead to immediate extinction.

The population dynamics is described by a balance equation for the
numerical gains and losses of each species.  This is written as
\begin{equation}
  \dot N_i=\lambda\sum_jg_{ij}N_i-\sum_jg_{ji}N_j-dN_i,
  \label{eq:N}
\end{equation}
where the three terms on the right-hand side correspond respectively
to gains from foraging, losses to predation, and losses to natural
death.  The factor $\lambda$ between losses to a prey species and
gains to its predator reflects the ecological efficiency of the
system, and we adopt a value of $\lambda=0.1$ consistent with
empirical data \citep[e.g.][]{pim82}.  We assign the scaling factor of
natural death, $d$, to be unity for all species, thus fixing the
time-scale of the model.  The remaining term, $g_{ij}$, comprises the
functional response.  We adopt a ratio-dependent functional response
that relates to a foraging strategy; $f_{ij}$ is the fractional effort
species $i$ puts into potential prey $j$, where $\sum_j f_{ij}=1$.
The functional response is given by
\begin{equation}
  g_{ij}=\frac{f_{ij}S_{ij}N_j}{bN_j+\sum_k\alpha_{ik}f_{kj}S_{kj}N_k};
  \label{eq:g}
\end{equation}
for very small predator populations this is approximately
\begin{equation}
  g_{ij}=\frac{f_{ij}S_{ij}}{b},
\end{equation}
so $b$ can be seen to restrict the feeding rate with high prey
availability.  A value of $b=0.005$ has been adopted from
\cite{dro01}, where it was found to give suitably realistic food webs.
The sum in the denominator reflects the effect of competition.
Competition is maximal between members of the same species, for which
$\alpha_{ii}=1$ for all $i$.  Competition with other species is
weaker, indicating the greater ability to extract resources from a
prey species given by a diversity of feeding techniques.  The formula
for $\alpha$ we also adopt from \cite{dro01} as
\begin{equation}
  \alpha_{ik}=c+(1-c)q_{ik},
  \label{eq:alpha}
\end{equation}
where $q_{ik}$ is the fraction of attributes shared by $i$ and $k$.
It can therefore be seen that $\alpha$ has a minimum value of $c$ for
two dissimilar species; we adopt the value of $c=0.5$, which has been
shown to produce reasonable food webs.

\cite{dro01} show that there exists an evolutionarily stable strategy
for $f$ given by
\begin{equation}
  f_{ij}=\frac{g_{ij}}{\sum_kg_{ik}},
  \label{eq:f}
\end{equation}
which we therefore take to yield the strategy for each species at all
times.

To maintain a limited size community, and in particular to give
meaning to the system size parameter $R$, species are removed from the
community if their population falls below unity.  From (\ref{eq:g}) it
can be seen that all values of $g_{ij}$ are unchanged if all
populations $N_i$ are changed by the same factor.  As such, if all
$\dot N_i$ are zero, (\ref{eq:N}) shows that this will still be the
case after scaling all $N_i$.  Hence all dynamics of the food web are
unchanged if both $R$ and the extinction threshold are scaled by the
same factor, and the arbitrary choice of unity population for
extinction is of no consequence.  In the absence of a non-zero
extinction threshold no extinctions ever occur, since populations will
at worst exponentially decay towards zero.

\subsection{Comparative models}
\begin{table}
  \begin{tabular}{lll}
    \hline\noalign{\smallskip}
    Name & Symbol & Definition \\[3pt]
    \vspace{1.5ex}number of species & $S$ &$\sum_i1$ \\
    \vspace{1.5ex}population & $N$ & $\sum_iN_i$ \\
    \vspace{1.5ex}abundance & $p_i$ & $N_i/\sum_jN_j$ \\
    \vspace{1.5ex}Shannon index & $H$ & $-\sum_ip_i\ln p_i$ \\
    \vspace{1.5ex}Simpson index & $D$ & $\sum_ip_i^2$ \\
    \vspace{1.5ex}Fisher index & $\alpha$ & $S=\alpha\ln\left(1+N/\alpha\right)$ \\
    \vspace{1.5ex}Shannon equitability & $E$ & $H/\ln S$ \\
    \vspace{1.5ex}trophic levels & $\Lambda$ & $\max_i\Lambda_i$ \\
    \vspace{1.5ex}mean species level & $\Lambda_S$ & $\sum_i\Lambda_i/S$ \\
    \vspace{1.5ex}mean individual level & $\Lambda_N$ & $\sum_ip_i\Lambda_i$ \\
    \vspace{1.5ex}links per species & $l$ & $\sum_il_i/S$ \\
    \noalign{\smallskip}\hline
  \end{tabular}
  \caption{
    Summary of community measures used in this paper.  Each species
    $i$ has population $N_i$, trophic level $\Lambda_i$, and $l_i$
    links to other species, counting both predators and prey.  A
    discussion of diversity measures can be found in \cite{mag88}, who
    denotes the Shannon index $H^\prime$.  Definitions of $H$
    sometimes use $\log_{10}$ or $\log_2$, and hence will differ in
    their results by a constant factor.
  }
  \centering
  \label{tab:measures}
\end{table}
The source community, from which species are drawn for immigration
into island communities, was built using the model described above,
with $R=10^6$. At the time at which the evolution run was stopped,
after $50\,000$ evolutionary events, 103 species were present, and the
food web was in a steady state configuration where the number of
species does not change systematically over long time intervals.  The
exact number of species would change on short timescales as evolution
and extinction events balance only on average.  This size of web was
chosen to give a reasonably large set of species for immigration
without taking excessive computational time to reach the steady state.

To construct an island community, the model was initialised with the
same matrix $m$ and environment species as the source community, but a
value of $R$ that was in general different.  Species were introduced
from the source community in a random order; the speciation mechanism
by which new species are introduced in the mainland model is not used,
so the number of species that can co-exist on any island is limited by
the number of species in the source community.  The parameter $R$ was
used as the independent variable to investigate the effect of resource
availability on the nature of the food web constructed.

For comparison with the effects of $R$ on the nature of the food web
constructed through immigration, mainlands were created for a range of
values of $R$ by the same process as the source community, using the
same matrix $m$ and environment species.  Different mainlands were
isolated in the sense that the species subject to population dynamics
were unrelated between mainland instances.

A second line of comparison comes from taking the source community and
performing population dynamics while gradually reducing $R$ from its
initial value.  There is no stochasticity in the result since the
population dynamics are deterministic and we are not introducing
species, so a single sequence of webs is produced.  We call this
sequence the reduction sequence, and for any value of $R$ refer to the
food web in the reduction sequence corresponding to resources $R$ as
the reduced food web of $R$.

\section{Results}
The results presented in this paper will cover only some of the
aspects of food web examination, and in particular will concentrate on
the variation with system size of measurements summarising food web
characteristics.  Equivalent figures are shown for measurements taken
from each of the mainland, island and reduction versions of the model.
Table~\ref{tab:measures} summarises the community measures used, where
$N_i$ is the population of the $i$th species, $\Lambda_i$ is its
trophic level, and $l_i$ is its total number of prey and predators.

\subsection{Island food webs}
\begin{figure}
  \centering
  \includegraphics[width=0.45\textwidth]{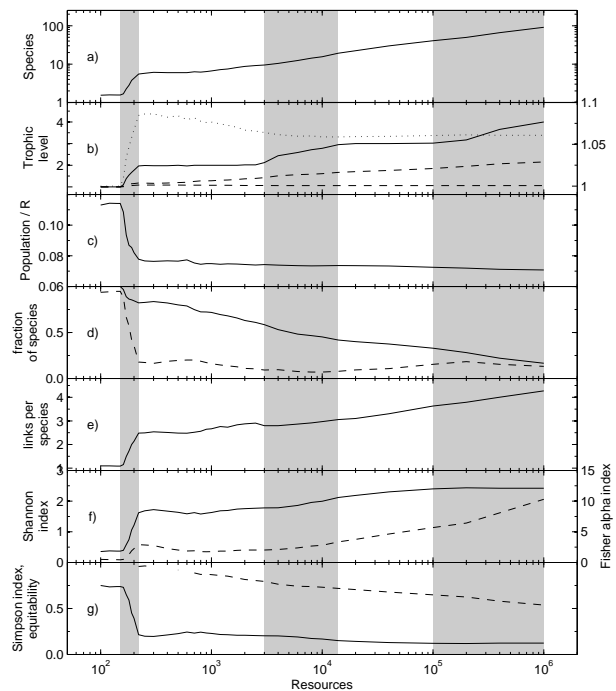}
  \caption{
    Food web measures taken from immigrant communities grown with
    various values of $R$.  The three vertical bands correspond to
    ranges of $R$ where the number of trophic levels changes,
    influencing other measures.  a)~Number of species present, $S$;
    b)~solid line~-- number of trophic levels, $\Lambda$; dashed
    lines~-- top, species average level, $\Lambda_S$, bottom,
    individual average level, $\Lambda_N$; dotted line~-- $\Lambda_N$
    against right hand axis; c)~Total population of all species, $N$,
    divided by $R$; d)~solid line~-- fraction of basal species; dashed
    line~-- fraction of unpredated species; e)~number of feeding links
    (in plus out) per species, $l$; f)~solid line~-- Shannon index,
    $H$; dashed line~-- Fisher's $\alpha$ index; g)~solid line~--
    Simpson's $D$ index; dashed line~-- equitability, $E$.
  }
  \label{fig:immigrate}
\end{figure}
In order to obtain meaningful averages over discretised quantities
such as the number of trophic levels in a food web, the results
plotted in figure~\ref{fig:immigrate} are the mean of measurements
taken from 100 simulation runs.  Continuously varying quantities, such
as the total population of all species, were found to rapidly converge
on this mean as the number of runs increased, implying a statistical
similarity of the food webs generated.  Results are shown for the
range of resources $100\le R\le 10^6$; the lower bound was chosen
because typically only one or two species were found in the resultant
food web, with very small total populations.  The behaviour of food
webs under formation as they approach the limit in which only one
species is supported, and that near extinction, is of little interest;
in any case a model using non-stochastic population dynamics is
inappropriate in this regime.  The upper limit of the resource range
is determined by the number of species in the source community; as
resources in the immigrant community increase, the number of species
present approaches that of the source, and the other measured
characteristics saturate also.  Saturation also occurs in time,
measured by the number of immigration events that have occurred, and
the samples examined were taken after 500 immigration events.  For the
largest values of resources there was some systematic change at later
times as the last few species were assimilated into the food web.

In figure~\ref{fig:immigrate}a) we plot the simplest measure of
success in assembling a food web from immigrant species, which is the
number of those species simultaneously present at the moment of
measurement.  As $R$ increases, almost all species are able to invade
the island, which suggests that the process of assembling a community
in this way is quite efficient.  A second simple feature to calculate
from a known food web is the number of trophic levels, plotted in
figure~\ref{fig:immigrate}b).  Here we use strictly defined trophic
levels equivalent to counting the number of feeding links in the
shortest route from each species to the environment on which all basal
species feed.  Thus all species feeding on the environment -- basal
species -- are in trophic level 1, species which feed on basal species
but not directly on the environment are in trophic level 2, and so
forth.  The number of trophic levels in any particular web is a
measurement of the highest single species, and thus is an integer.
The solid line in figure~\ref{fig:immigrate}b) shows the mean value of
this quantity over simulation runs, and hence is not necessarily
integer.  It can immediately be seen that the formation of food webs
with multiple trophic levels does occur when species are supplied by
immigration from a previously evolved food web.  We note that the
increase in the number of trophic levels with $R$ is not smooth, and
that for certain ranges of resources, essentially all webs have the
same number of trophic levels.  For example, in the range $200<R<2000$
the mean number of trophic levels is almost exactly 2, and a similar
plateau occurs at a mean trophic level of 3.  The transition from a
single trophic level to two is particularly sharp, with the onset of
higher trophic levels being spread over a larger range of resources.
Vertical bands in figure~\ref{fig:immigrate} mark the periods of
transition between trophic levels, and the first transition can be
seen to correspond to a rapid increase in the number of species
present.  Examination of individual food webs suggests that this
occurs when `herbivores', which feed on the basal `plants', enter the
system.  A herbivorous species reduces the success of the plants on
which it feeds, which in turn allows other plants to compete for
resources effectively.  The Webworld model thus displays
predator-mediated coexistence.  For slightly larger $R$
figure~\ref{fig:immigrate}a) shows a plateau in the number of species
present before an approximately power law increase in
$S\!\left(R\right)$ sets in.

Figure~\ref{fig:immigrate}b) also plots, as dashed lines, the average
trophic level.  The upper dashed line indicates the mean trophic level
of a species in the food web, which increases more smoothly than the
number of trophic levels.  A new trophic level is founded by a single
species; as the population of the species on which it feeds increase,
it becomes possible to support a greater number of species on the top
level, and hence the mean trophic level increases while the number of
distinct levels is constant.  The lower dashed line indicates the mean
trophic level of individuals in the food web, and is lower than the
mean by species since the population of species tends to decrease with
increasing trophic level.  To examine this curve more closely it has
been re-plotted as the dotted line against the right-hand axis.  Here
it can be seen that there is a maximum in this quantity immediately
after the onset of the second trophic level.  Thereafter the
population of basal species increases more rapidly than for other
trophic levels, and $\Lambda_N$ decreases asymptotically.  This is
related to the result plotted in figure~\ref{fig:immigrate}c), which
shows that after the onset of the second trophic level, the number of
individuals summed over all species is nearly a constant fraction of
$R$.  Considering the case of a single species feeding on the
environment, we can identify its population by setting the time
derivative in (\ref{eq:N}) to zero and substituting $f=1$ in
(\ref{eq:g}).  For a given value of $S$, this can be used to deduce
the population of the species to be
\begin{equation}
  \frac{N_{\rm basal}}R=\lambda\left\{\frac1d-\frac{b}S\right\},
\end{equation}
and hence cannot exceed 0.1 for our value of $\lambda$.  The larger
value of $N/R$ seen at the left of figure~\ref{fig:immigrate}c)
corresponds to the presence of multiple basal species, where the
reduced inter-specific competition allows a larger total population.
The steady value of $N/R$ for multiple-trophic level webs seems to be
an effect of food web regulation.

Figure~\ref{fig:immigrate}d) shows the fraction of species that are
basal (solid line), and the fraction that are unpredated (dashed
line).  The fraction of basal species falls steadily as resources
increase and more species are able to find niches at higher trophic
levels.  The fraction of top species, i.e. those that are unpredated,
has a remarkably different behaviour, dropping to an almost constant
level as soon as trophic structure exists.  As new species find a
niche in the food web they have small population and are likely to be
unpredated, but at the same time typically find that niche by feeding
on a species that was previously unpredated.  As $R$ increases, all
species present typically increase in population, and at some point a
top species becomes an exploitable prey.  The fluctuations in the
fraction of top species, $T$, suggest this quantity emerges from food
web regulation, and we do not expect a simple explanation for
structure in $T$ as a function of $R$.

\begin{figure}
  \centering
  \includegraphics[width=0.45\textwidth]{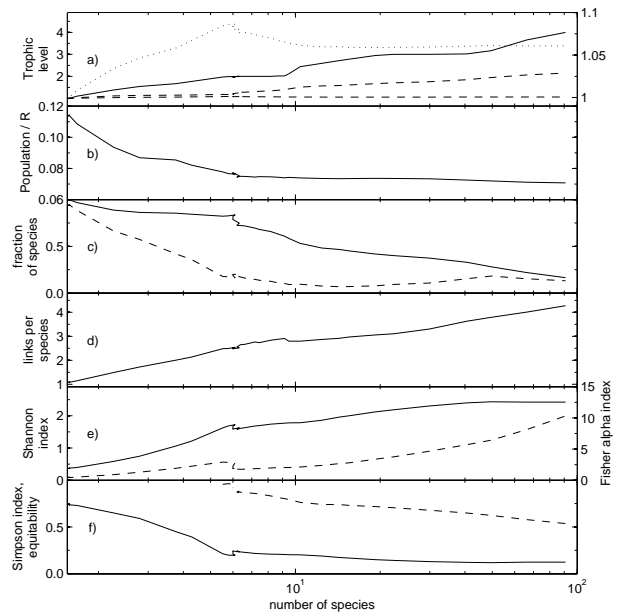}
  \caption{
    Food web measures taken from immigrant communities.  The same data
    plotted in figure~\ref{fig:immigrate} are here plotted against
    number of species, $S$.  Figure~\ref{fig:immigrate}a) plots $S$ as
    the dependent variable, and is therefore omitted, leaving;
    a)~solid line~-- number of trophic levels, $\Lambda$; dashed
    lines~-- top, species average level, $\Lambda_S$, bottom,
    individual average level, $\Lambda_N$; dotted line~-- $\Lambda_N$
    against right hand axis; b)~Total population of all species, $N$,
    divided by $R$; c)~solid line~-- fraction of basal species; dashed
    line~-- fraction of unpredated species; d)~number of feeding links
    (in plus out) per species, $l$; e)~solid line~-- Shannon index,
    $H$; dashed line~-- Fisher's $\alpha$ index; f)~solid line~--
    Simpson's $D$ index; dashed line~-- equitability, $E$.
  }
  \label{fig:immigrateS}
\end{figure}
Figure~\ref{fig:immigrate}e) indicates that the complexity of the food
web increases with increasing resources in terms of the number of
feeding links associated with each species.  To better interpret the
structure in such plots we present, in figure~\ref{fig:immigrateS},
the same plots as in figure~\ref{fig:immigrate} but plotted against
$S$.  Features in the previously discussed quantities seem better
understood in terms of resource availability, but
figure~\ref{fig:immigrateS}d) suggests that the food webs produced by
this model have a typical connectance increasing almost proportionally
to $\log S$.  The connectance of these food webs decreases with $S$,
but on average species become more connected.

The remaining plots, figure~\ref{fig:immigrate}f) and g),
corresponding to figure~\ref{fig:immigrateS}e) and f) respectively,
show measures of diversity in the food web.  Diversity indices each
summarise the species abundance distribution (SAD) of an ecosystem as
a single number, and \cite{pla84} note that different indices will not
provide consistent ranking among ecosystems because they each measure
a different aspect of diversity.  Where one food web is unambiguously
more diverse than another we expect to see similar trends in each of
the diversity indices, but differences between them are also expected,
reflecting the changing SAD.  The first diversity index we show is the
commonly used Shannon index of the system, with values here based on
the natural logarithm.  Typical values of this quantity for natural
ecosystems are suggested to be in the range $1.5-3.5$ \citep{mag88},
while the range measured from our simulations is $1.5<H<2.5$,
indicating a lower diversity and/or equitability.  For comparison we
calculate $H$ from the data set for tree abundances on Barro Colorado
Island (BCI), Panama in 2005 \citep{hub05}.  For the full set of 299
tree species with non-zero population, the Shannon index has a value
of nearly 4.  To match $H$ as determined from our simulation results,
the abundance distribution must be truncated after 35 species. An
important difference between the BCI data and our simulation results
is that the former comprises a single trophic level.  The food supply
available to a `herbivore' in our model is approximately the total
population of all species, which figure~\ref{fig:immigrate}c) shows to
be approximately $0.08R$.  We expect the herbivore population to be
reduced by a corresponding factor.  To model this, we sort the BCI
data by abundance, and except for the first $x$ species reduce the
abundance of each species by a factor of $0.08$.  For $x\simeq10$ we
obtain $H \simeq2.5$.  This is similar to our simulation results with
a similar number of basal species.  The sudden increase in $H$, and
drop in the Simpson index $D$, during the formation of the second
trophic level correspond to the increasing number of basal species
which partition the total basal population.  Further species additions
tend to occur at higher trophic levels, and hence have a much smaller
population, contributing diminishing amounts to the Shannon index as
the total number of species increases. The form of the species
abundance distribution (SAD), which is summarised by the diversity
indices, will be examined in a future paper.

Fisher's $\alpha$ diversity index \citep[e.g.][]{mag88} is often
favoured as a diversity index which does not depend on sample size
\citep[e.g.][]{ros95}, being unaltered as increasingly rare species
are sampled from an underlying distribution.  The fact that $\alpha$
increases as a function of $R$ therefore indicates that the underlying
distribution is changing; it is not an adequate description to suppose
that the species present are those whose population in some static
distribution lies above a threshold which decreases with increasing
$R$.

\subsection{Mainland food webs}
\begin{figure}
  \centering
  \includegraphics[width=0.45\textwidth]{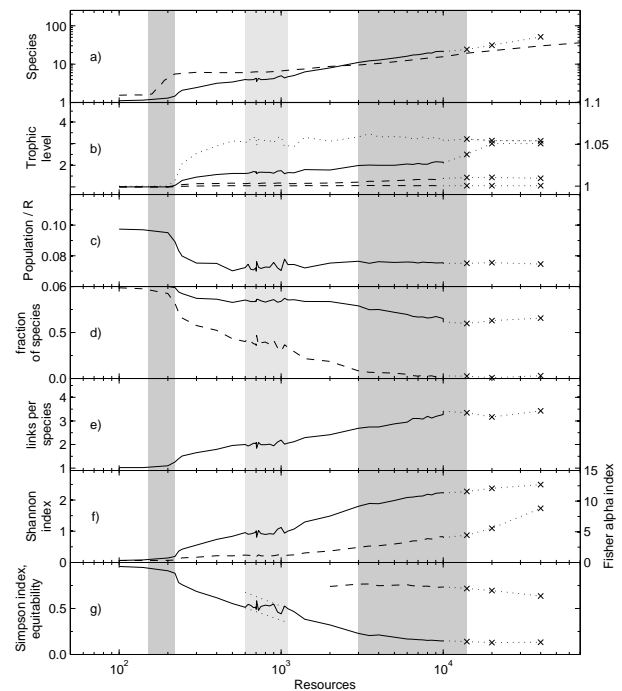}
  \caption{
    Food web measures taken from mainland communities grown with
    various values of $R$.  The outer vertical bands correspond to the
    first two bands of figure~\ref{fig:immigrate}; the middle band is
    described in the text.  a--g) as figure~\ref{fig:immigrate}.
    Marked points connected by dotted lines correspond to averages
    over ten runs; other data are averaged over 100 runs.
  }
  \label{fig:evolve}
\end{figure}
Whereas island communities reach a steady state after only a few
hundred immigration events, tens of thousands of evolution events are
required to reach the corresponding steady state; here we use results
from runs of length $50\,000$ evolutions.  As such the computational
demands of mainland ecosystems are far greater than those of islands,
and we have not been able to calculate statistical results for such a
wide range of resources.  A further factor increasing computational
complexity is the fact that as $R$ becomes large there are more
species on mainlands than on islands, where $S$ cannot exceed that of
the source community.  Averages in figure~\ref{fig:evolve} are over
one hundred runs for the solid lines, as with the island results, but
are averages over only ten runs for the dotted lines corresponding to
larger $R$.  It was noticed that the results of averaging over even
one hundred simulation runs were particularly poor in the range of
resources marked by the central vertical band in
figure~\ref{fig:evolve}, which manifests itself as a large variation
in most quantities plotted within this range.  This effect can be seen
most clearly in figure~\ref{fig:evolve}g), the Simpson index $D$,
which appears to be in transition between two values marked by dotted
lines.  The main cause of these fluctuations is related to the number
of webs possessing different number of species, with $S$ for an
individual mainland web in the range $1\le S\lsim10$ for $R=700$.

\begin{figure}
  \centering
  \includegraphics[width=0.45\textwidth]{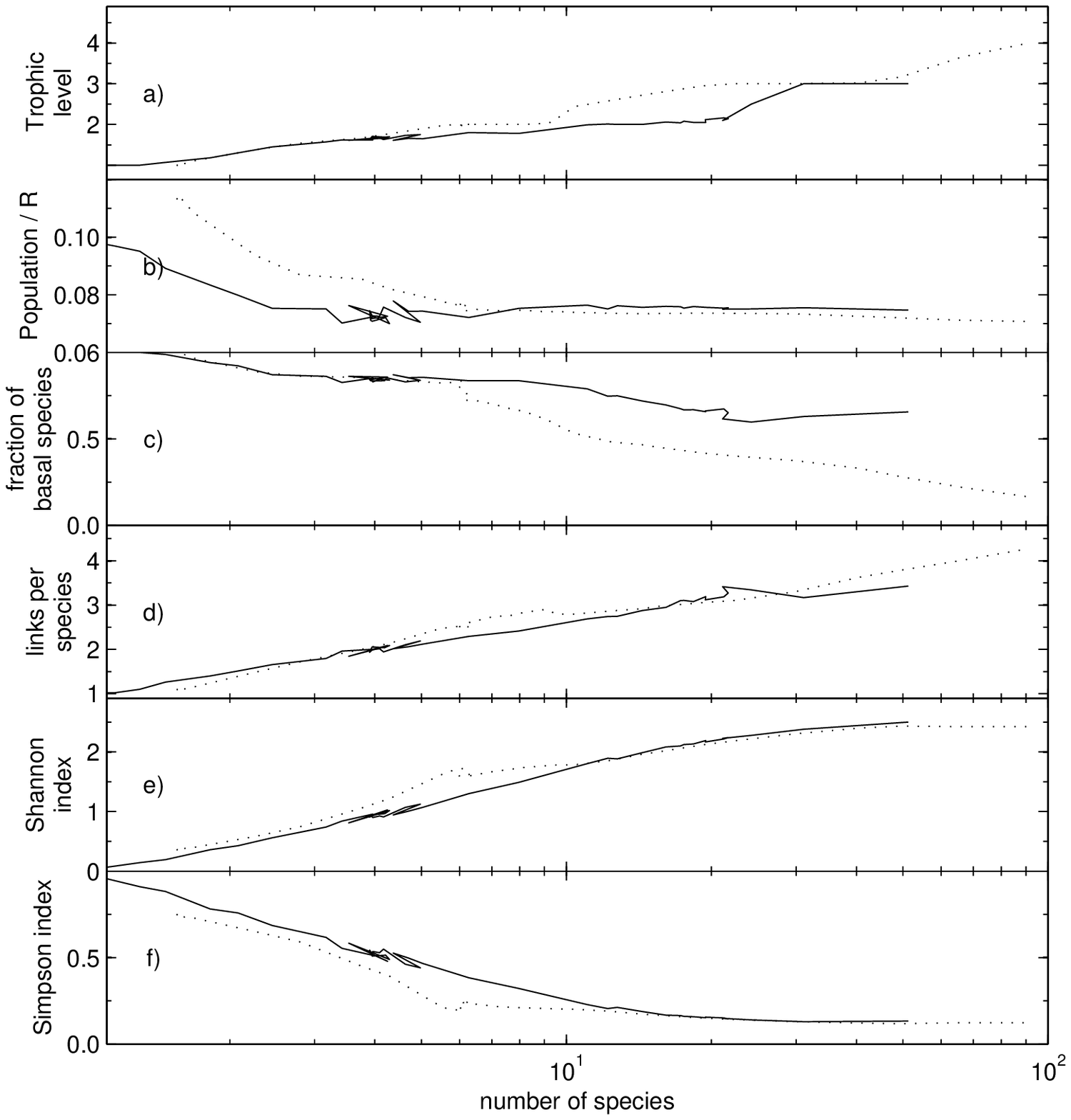}
  \caption{
    Comparison of selected quantities between island and mainland food
    webs, as a function of $S$.  Solid lines indicate mainland data
    and dotted lines island data repeated from
    figure~\ref{fig:immigrateS}.  a)~number of trophic levels,
    $\Lambda$; b)~Total population of all species, $N$, divided by
    $R$; c)~fraction of basal species; d)~number of feeding links per
    species, $l$; e)~Shannon index, $H$; f)~Simpson's $D$ index.
  }
  \label{fig:evolveS}
\end{figure}

The most obvious difference between the results for a food web grown
through evolution when compared to the island food webs is that there
are no longer distinct boundaries corresponding to the onset of
trophic levels.  When plotted against $S$ in
figure~\ref{fig:evolveS}a) there appears to be a relatively sharp
onset of the third trophic level, but these data are averages over
only ten realisations, and may be unreliable.  Compared with island
food webs, the onset of trophic levels is at larger $R$ and, as shown
in figure~\ref{fig:evolveS}a), also larger $S$.  The left and right
vertical bands in figure~\ref{fig:evolve} mark the onset of the second
and third trophic levels respectively in island food webs, and
mainland food webs do not begin to attain those levels until the
right-hand edge of the corresponding band.  We hypothesise that for
small $R$, where island communities support more species than
mainlands, the effect of species adapting to their predators and prey
makes it more difficult to attain high trophic levels.  Prey species
in island communities cannot adapt to evade predators in our model, so
predators can succeed more easily.

As remarked, for $R\lsim3000$ an island community supports more
species than a mainland; the dashed line in figure~\ref{fig:evolve}a)
repeats the curve for islands shown in figure~\ref{fig:immigrate}a).
As species evolve to become better at feeding on their prey and
avoiding their predators, they tend to reduce the population of those
species until they in turn evolve.  As such, for small $R$ it is
easier for one species to eliminate another from the food web, and the
steady state number of species is smaller.  As $R$ increases, the
availability of new species to an island diminishes, whereas for a
mainland the amount of novelty possible increases with $S$ leading to
a potentially unbounded value of $S$ as $R\rightarrow\infty$.
Certainly the total number of species possible in the Webworld model
is far larger at $\sim10^{20}$ (for the particular parameter values we
have adopted here) than the number simultaneously present in any web
created, and the dotted line in figure~\ref{fig:evolve}a) appears to
continue the trend of an approximately power law increase in $S$ with
$R$, $S\propto R^{0.63}$.

A striking difference between island and mainland food webs appears in
the fraction of top species, shown in figure~\ref{fig:evolve}d) as the
dashed line.  For island food webs the fraction of top species is
small and nearly constant for food webs of multiple trophic levels,
whereas for mainland communities it decreases gradually toward zero
from a higher starting point.  This probably reflects the potential
for new species to enter the food web specialising in previously
unpredated species, while in the island community species are
constrained by their existing abilities.  The fraction of basal
species, shown as the solid line in the same plot, is somewhat higher
in the mainland community, but with a similar trend to decrease with
$R$.  Examination of figure~\ref{fig:evolveS}c) shows that for
$S\lsim6$ island and mainland communities behave similarly, but for
larger $S$ the fraction of basal species in island communities starts
dropping more rapidly.  This is not obviously associated with the
number of trophic levels, which is not seen to increase significantly
in figure~\ref{fig:evolveS}a) for island food webs at $S\simeq6$.
Conversely it is at about $S=6$ that the total population of island
and mainland communities as a fraction of resources, shown in
figure~\ref{fig:evolveS}c), converge.  For smaller $S$ island
communities have a higher total population, and both settle to a
similar value for larger $S$, with islands having slightly lower
abundance.

In comparing the number of links per species between island and
mainland food webs, it is found that the differences shown in
figure~\ref{fig:evolveS}d) are neither large nor systematic.  As such
it seems that the number of feeding links a species possesses is
primarily dictated by the number of species in the ecosystem as a
whole, with relatively minor contributions from probable influences
such as the trophic structure and direct consequences of assembly.
Diversity indices $H$ and $D$, shown in figures~\ref{fig:evolveS}e)
and \ref{fig:evolveS}f) respectively, also show a behaviour that is
largely similar for island and mainland communities.  There is an
initial increase in diversity which is nearly linear in terms of $H$
versus $\log S$, turning down for $S\gsim20$.  This later decrease in
the derivative is probably related to the fact that species are
increasingly added at higher trophic levels where the population is
automatically smaller; $H$ is maximised by a uniform distribution.
The strongest difference between the diversity of island and mainland
communities occurs at $S\simeq6$, where there is an anomalous excess
in $H$ for islands.  The maximum difference coincides with the sharply
emerging difference in the fraction of basal species shown in
figure~\ref{fig:evolveS}c), and is presumably related.  For $H$ to be
higher in island communities when most species are basal indicates
that there is a more even distribution of population between those
basal species.  The rapid increase in the fraction of species on the
second trophic level diminishes $H$ by the previously noted effect of
population decreasing with level, restoring equality between island
and mainland food webs.

\subsection{Reduction food web}
\begin{figure}
  \centering
 \includegraphics[width=0.45\textwidth]{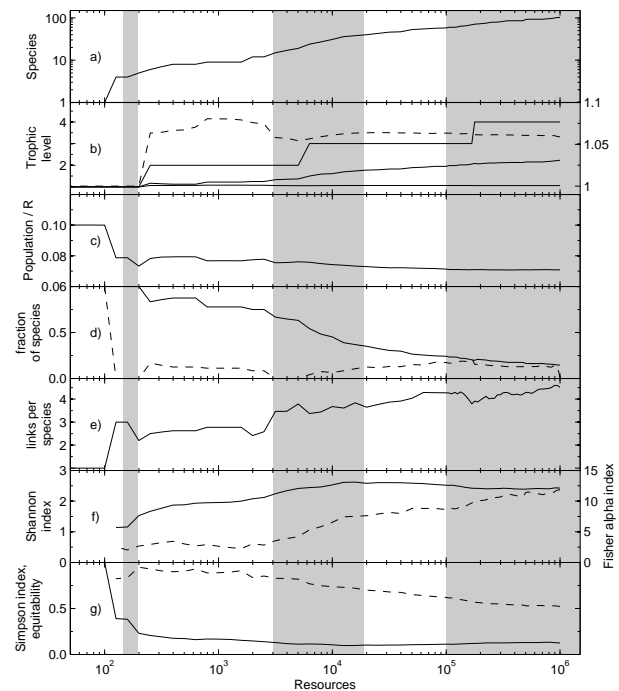}
  \caption{
    Food web measures taken as the source community was run with
    reducing $R$.  Vertical bands correspond to those of
    figure~\ref{fig:immigrate}.  a--g) as figure~\ref{fig:immigrate}.
  }
  \label{fig:reduce}
\end{figure}

\begin{figure}
  \centering
  \includegraphics[width=0.45\textwidth]{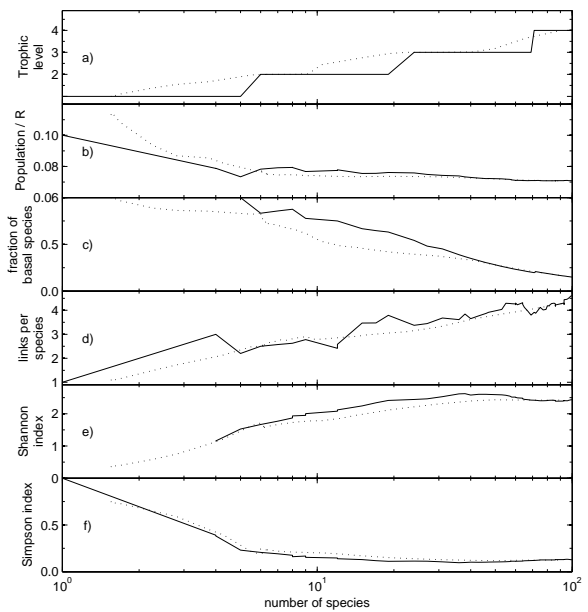}
  \caption{
    Comparison of selected quantities between island food webs and the
    reduction sequence, as a function of $S$.  Solid lines indicate
    reduction data and dotted lines island data repeated from
    figure~\ref{fig:immigrateS}.  a--f) as figure~\ref{fig:evolveS}.
  }
  \label{fig:reduceS}
\end{figure}

The number of species in the single reduction process from the island
source community necessarily starts at 103 for $R=10^6$, corresponding
to the source community itself.  For this value of resources the
island community does not generally reach the full compliment of
species, and the mean value is considerably lower.  As shown in
figure~\ref{fig:reduce}a), as resources decrease species drop out of
the reducing food web gradually rather than triggering catastrophic
losses, and the number of species in the reducing food web is larger
than the number in the immigrant community for all $R$ despite the
limitation that species cannot be reintroduced.  The limitation that
we have only one possible history to examine means that there is an
integer number of trophic levels for all $R$ in the reduction model
(non-integer values plotted in figure~\ref{fig:reduce}b) lie between
sample points), so plateaux are an inevitability.  These plateaux
roughly correspond to the ranges for which plateaux are seen in the
corresponding results for island food webs, shown in
figure~\ref{fig:immigrate}b).  Figure~\ref{fig:reduceS}a) shows that
the island communities tend to have more trophic levels for the number
of species than the reduction model, although as mentioned they have
fewer species for a given $R$.

In general the trends seen in the reduction model are very similar to
those seen in the island food webs.  The fraction of basal species in
the island and reduction webs approach each other for large $S$; for
small $S$ this fraction tends to unity more rapidly as $R$ decreases
for the reduction model, possibly reflecting a tendency to
preferentially remove non-basal species.  The difference seems to
relate to the trophic structure shown in figure~\ref{fig:reduceS}a),
and might be removed if reintroduction of species were allowed in the
reduction model.  Measurements of $N$ and $l$ shown in
figures~\ref{fig:reduceS}b) and \ref{fig:reduceS}d) respectively are
less smooth for the reduction model due to the absence of averaging,
but are otherwise consistent.  The small increase in the Shannon index
of the reduction model over islands, shown in
figure~\ref{fig:reduceS}e), is probably related to a preference for
removing species of higher trophic level, thus making the average
species more like a basal species in population and increasing
uniformity.  This increase in diversity is also shown in the Simpson
index in figure~\ref{fig:reduceS}f).

\section{Species-area relation}
One of the most studied features related to the effect of size on food
web structure is the species-area relation (SAR), because this is
relatively easy to sample for real systems.  We model the SAR by
equating area to resources, $R$, with which we supply basal species,
and examine the equilibrium number of species for food webs grown with
different values of $R$.  Several distinct types of SAR can be formed
depending on the sampling method; that applicable to our data was
labelled as the Type IV species-area curve by \cite{sch03},
corresponding to sampling of physically distinct regions (islands or
other isolated units) of differing area.  \cite{gra04a} prefer to
reserve the term ``species-area curve'' for just this case, but see
\cite{sch04} and \cite{gra04b} for further discussion.  Although a
monotonic increase in $S$ with area is intuitive for any SAR, it is
not a necessary consequence of the Type IV sampling method; large
islands may in principle have fewer species than smaller islands.

Some of the effects found in real ecosystems which influence the shape
of the SAR are not applicable to the data we collect on Webworld.  In
particular, \cite{ros95} discusses three reasons to expect more
species in a larger sample area.  Firstly, a larger sample area may
imply a larger number of collected individuals, giving a more complete
sample.  For Webworld we have complete population information at any
time in the simulation, and in particular can distinguish with
accuracy whether a species is present or absent.  The second effect
relates to diversity of habitats; a larger area in a physical system
is likely to contain heterogeneous regions to which different species
are better adapted.  The Webworld model has only a single habitat type
independent of the `area' we attribute to that system, and additional
niches only arise through the existence of exploitable populations of
other species.  The third factor discussed in complicating SARs
measured from real data is the existence of biogeographical provinces;
a very large system may contain areas which could be occupied by the
same set of species except for historical exclusion.  Examples abound
of apparently identical niches being filled by different species in
isolated regions.  Since the Webworld model we use is only implicitly
spatial, no such geographical barriers exist to maintain diversity,
and this explanation for an increase in species count with area can
also be excluded.

\cite{ros95} contests the findings of \cite{bos83} that the number of
species increases with area for a homogeneous environment, indicating
that Fisher's diversity index, $\alpha$, does not systematically
depend on area for their samples.  As shown in
figures~\ref{fig:immigrate}f), \ref{fig:evolve}f) and
\ref{fig:reduce}f), our results show a significant trend in $\alpha$
with area.  We conclude that in our results the increase in species
number with area is not merely due to changes in the cut-off
population of otherwise similar distributions.  For the smallest food
webs this is particularly obvious in the food web structure, since the
coexistence of basal species (`plants') is strongly promoted by the
presence of a single `herbivorous' species.

The method of immigration we have used in the context of the Webworld
model is similar to the premise of the theory of island biogeography
by \cite{mac63}.  However, the population dynamics in Webworld is
deterministic, so between the introduction of species the population
of each species settles to a steady value, and hence the rate of
extinction becomes zero.  As such, extinctions are related to the
displacement of species by new arrivals rather than through
fluctuations per se.  It is still the case that those species driven
extinct are those unable to adopt a feeding strategy that will support
an equilibrium population above our extinction threshold.

\begin{figure}
  \centering
  \includegraphics[width=0.45\textwidth]{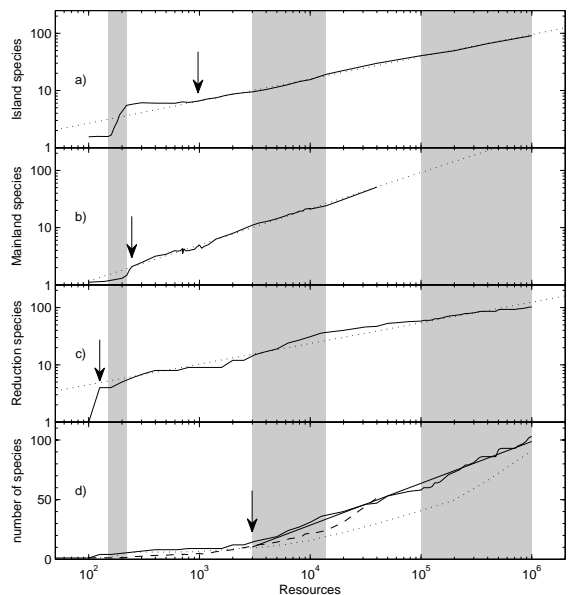}
  \caption{
    Species-area relations for the three model variants; a)
    immigration, b) evolution and c) reduction.  Dotted lines mark
    power-law fits to data to the right of the corresponding arrow.
    Power law indices are 0.39, 0.63, and 0.36 respectively.  d)
    Gleason plot of the above data.  Solid line -- reduction data and
    fit of data right of the arrow ($S=35\log R-112$); dotted line
    immigration; dashed line evolution.
  }
  \label{fig:sar}
\end{figure}

In figure~\ref{fig:sar} we show log-log plots of the SAR of islands,
mainlands and the reduction sequence in \ref{fig:sar}a), b) and c)
respectively; in \ref{fig:sar}d) we plot for comparison against a
linear $y$-axis (Gleason plot).  In this case it can be seen that only
for the reduction sequence is a linear regression a reasonable
approximation across any of the range.  For at least the island and
mainland models, a power law fit to the data is acceptable for all $R$
greater than some threshold, indicated by arrows in
figure~\ref{fig:sar}.  The behaviour of $S$ with area for small $R$ is
discussed in the results section.

Power law indices for the immigration and reduction models are
similar, whereas the evolution model has a rather steeper power law.
\cite{ros95} attributes the degree of fitting of observed SARs to the
form $S=k_1+k_2\log A$ for constants $k_1$, $k_2$ and area $A$ to
sampling effects in small food webs; \cite{ryd88} note that the SAR
including large islands is only well fitted by a power law, despite
their conclusion that a logarithmic increase is a better description
overall.

The shape of the SAR identified by \cite{pre60} is tri-phasic; $z$ is
larger for both small and large areas than for the range of area
between.  Although we see anomalous behaviour for small areas in the
Webworld results, we cannot associate a power law with these results
since we do not see a straight line in the log-log plot.  \cite{dur96}
collect examples of typical power law indices in their Table 1, which
have the approximate range $0.15\le z\le 0.5$, consistent with our
results for the island and reduction models, which have $z=0.39$ and
$z=0.36$ respectively.  The power law index we find for the mainland
model exceeds this range somewhat, at $z=0.63$.  \cite{hub01}
describes the part of the tri-phasic curve corresponding to the
largest areas as the `continental' scale, where diversity is
attributable to evolutionary processes.  Values for $z$ as high as
unity are possible for intercontinental scales; $z=1$ would be found
if two sample areas of equal size and no overlap in species were
considered.  \cite{ros95} cites examples of data from which
species-area curves with $z=1$ can be extracted, corresponding to
large length scales.

A different context within which a high power law index $z$ is found
for the species-area relation was identified by \cite{car06}, who
study the changing SAR during stages of succession on Mount St. Helens
following volcanic sterilisation.  \cite{car06} found that the power
law index of the SAR decreased with time as species became
re-established, and in their figure~2 present the SAR of Abraham
Plains for 1988 in which $z=0.69$.  We hypothesise that the large
value of $z$ we find for mainland data may be due to internal
disruption caused to the food web structure by the overturn of
species, noting that in small food webs one effect of this overturn is
to remove specialised predators whose prey becomes replaced.  A system
in which evolution and immigration were both allowed to play a part
would be expected to display a power law index intermediate between
those found for our immigration and evolution models, and be
consistent with the concept of lessened disruption.

\section{Discussion and conclusions}
In this paper we make use of the Webworld model introduced by
\cite{cal98} and \cite{dro01} to examine food web assembly through
immigration.  Previous work on this topic, in particular assembly
models \citep{law99}, have drawn immigrant species from a randomly
assigned species pool, and as such their results may not reflect the
properties of a natural ecosystem.  Instead we create the species pool
itself by evolution, and draw species from this stable community to
construct `island' food webs, where in this discussion we use the term
`island' to relate to ecosystems constructed solely by immigration.
Food webs constructed solely through evolution in situ we designate
`mainland'.  In order to compare our results with measurements of real
ecosystems we equate the resources supplying the Webworld model to
physical area, noting that not being explicitly spatial, Webworld can
only represent systems without geographical barriers.

It is of interest to study food webs assembled through immigration not
only because these correspond to most of the real world ecosystems for
which data is available, and especially to any experimental set-up
which could be created, but also because from a modelling point of
view the assembly of food webs through immigration is much more
repeatable than the evolutionary scenario, in which the identity of
the species is necessarily a chance occurrence.  For each food web
grown through evolution we can in shorter time examine the statistical
properties of the family of island communities which can be grown from
it.

One of the clearest differences between island and mainland food webs
is the power law index of the species-area relation.  For our islands
we find this to be, at $z=0.39$, within the range found for natural
ecosystems, typically $0.2\le z\le 0.4$.  For our mainland results we
find the steeper power law of $0.63$, and note that a steeper power
law in nature is associated with the very largest length scales, where
evolution is more important than immigration due to the presence of
substantial geographical barriers.  It is difficult to obtain a
species-area relation for real ecosystems in which only evolution
played a role because these would need to correspond to regions
sufficiently small to promote complete mixing and sufficiently remote
to avoid immigration from a continent, yet numerous enough to provide
good statistics for essentially similar environments.  We note that a
high power law may also be associated with a strongly perturbed
system, e.g. \cite{car06}, and that the continual turnover of species
through evolution may provide such a disturbance.

A second significant difference between island and mainland food webs
is found in the fraction of basal and of top species.  For island food
webs the fraction of top (unpredated) species does not change
systematically over a large range of system size, or equivalently as a
function of the number of species.  For mainlands with a large number
of species ($S>10$) there are far fewer top species than on islands.
Conversely islands have a smaller fraction of basal species than
mainlands in our results.  Considering the progression of food webs
from small area to large, we observe that islands show distinct
plateaux in the number of trophic levels, with all islands having the
same number of levels for large ranges of area.  Mainland food webs
may show a weaker tendency toward this phenomenon, but we cannot
exclude a smooth increase in the mean number of levels with area
across realisations.

Despite these important differences, much of the behaviour of the
model is essentially similar whether immigration or evolution supplies
the species.  In particular the number of trophic levels is close for
islands and mainlands of similar area, and the number of feeding links
per species is also very similar, implying statistical similarities
between the food webs constructed.  Such similarities are reflected in
the Shannon index of the system, which increases nearly linearly in
$S$ for small food webs, with a decrease in the gradient for $S>10$.
The Shannon index is associated with the distribution of population
sizes, which we will examine in a future paper.

Several avenues of future work are suggested by the results we have
found.  We have performed preliminary work in which immigration takes
place along a chain of islands, with immigration to one island only
being possible from its predecessor in the chain.  Our results to date
suggest that this induces changes in the nature of the food webs
several links down the chain, with a `filtering' effect precluding the
transmission of species which can be easily displaced.  Similar
scenarios reminiscent of real archipelagos may provide interesting
comparison with real systems.

Another change to bring our model more into line with real systems
would be to interpolate between the pure evolution and pure
immigration models by allowing both to take place concurrently to
different degrees.  Of particular relevance is the case in which a
small degree of evolution is allowed, to determine whether this has a
significant effect on the island properties we have identified in this
paper.

All of the islands for which results are presented in this paper were
grown from a single source community.  Beside confirming that the
results are not significantly altered depending on the course
community used, we are in principle able to examine the behaviour of
an island where immigration occurs from two or more independent
mainlands.  A related complexification of the model which may relate
to real ecosystems is to introduce multiple environment species to
simulate the presence of different habitats.  This work may require a
spatially explicit model for which computational resources are not yet
available.

Finally we note that one of the most important food web
characteristics which we have not examined is the species abundance
distribution (SAD).  This is related to the various diversity measures
in that they provide different scalar measures of this distribution.
The species-area relation is also related to the dependence of the SAD
on area.  The examination of the SAD requires substantial detail, and
our results for this will be presented in a future paper.

\section*{Acknowledgements}
We wish to thank the EPSRC (UK) for funding under grant number GR/T11784.

\end{document}